\documentclass[aps,pra,nofootinbib,twocolumn]{revtex4}
\usepackage{amsmath}
\usepackage{amssymb}
\usepackage{epsfig}
\usepackage{bm}

\def\be{\begin{equation}}
\def\ee{\end{equation}}
\def\bea{\begin{eqnarray}}
\def\eea{\end{eqnarray}}

\begin{document}

\title{Simplified equations for gravitational field in the vector theory of
gravity and new insights into dark energy}
\author{Anatoly A. Svidzinsky}
\affiliation{Department of Physics \& Astronomy, Texas A\&M University, College Station
TX 77843-4242}
\date{\today }

\begin{abstract}
Recently, a new alternative vector theory of gravity has been proposed which
assumes that universe has fixed background Euclidean geometry and gravity is
a vector field that alters this geometry [Phys. Scr. 92, 125001 (2017)]. It
has been shown that vector gravity passes all available gravitational tests
and yields, with no free parameters, the value of cosmological constant in
agreement with observations. Here we obtain substantially simplified
gravitational field equations of vector gravity which are more suitable for
analytical and numerical analyses. We also provide a detailed explanation
why in vector gravity in the reference frame of observer that takes a
snapshot of the universe at time $t_0$ the ratio of the cosmological
constant to the critical density is equal to $2/3$ at $t=t_0$. We also show
that dark energy does not affect universe evolution in the co-evolving
reference frame. Thus, in reality, universe is expanding at a continually
decelerating rate, with expansion asymptotically approaching zero.
\end{abstract}

\keywords{vector gravity; dark energy; cosmological constant}
\maketitle

\section{Introduction}

Recently, a new alternative vector theory of gravity has been proposed \cite%
{Svid17}. The theory assumes that gravity is a vector field in fixed
four-dimensional Euclidean space $\delta _{ik}$ which breaks the original
Euclidean symmetry of the universe and alters the space geometry. The
direction of the vector gravitational field gives the time coordinate, while
perpendicular directions are spatial coordinates.

Conceptually, both vector gravity and Einstein's general relativity deal
with the space-time geometry. However, in general relativity the space-time
geometry itself is the (dynamic) gravitational field. The concept of vector
gravity is that the fixed background geometry is altered by the vector
field. Similar concept applies to any metric theory of gravity with a prior
geometry \cite{Will93}.

Despite fundamental differences from general relativity, it has been shown
that vector gravity passes all available gravitational tests \cite{Svid17}.
In particular, vector gravity and general relativity are equivalent in the
post-Newtonian limit and, thus, they both pass solar-system tests of
gravity. The two theories also give the same quadrupole formula for the rate
of energy loss by orbiting binary stars due to emission of gravitational
waves.

In strong fields, vector gravity deviates substantially from general
relativity and yields no black holes. Since the theory predicts no event
horizons, the end point of a gravitational collapse is not a point
singularity but rather a stable star with a reduced mass. In vector gravity,
neutron stars can have substantially larger masses than in general
relativity and gravitational wave detection events can be interpreted as
produced by the inspiral of two neutron stars rather than black holes \cite%
{Svid17}. Vector gravity predicts that the upper mass limit for a
nonrotating neutron star with a realistic equation of state is of the order
of $35$ M$_{\odot }$ (see Sec. $13$ in \cite{Svid17}). Stellar rotation can
increase this limit to values in the range of $50$ M$_{\odot }$. The
predicted limit is consistent with masses of compact objects discovered in
X-ray binaries \cite{Casa14} and those obtained from gravitational wave
detections \cite{LIGO18}.

Vector gravity also predicts that compact objects with masses greater than $%
10^{5}M_{\odot }$ found in galactic centers have a non-baryonic origin and
made of dark matter. It is interesting to note that their properties can be
explained quantitatively in the framework of vector gravity assuming they
are made of dark matter axions and the axion mass is about $0.6$ meV (see
Sec. 15 in \cite{Svid17} and Ref. \cite{Svid07}). Namely, supermassive
objects at galactic centers are axion bubbles. The bubble mass is
concentrated in a thin surface - the interface between two degenerate vacuum
states of the axion field. If the bubble radius is large, surface tension
tends to contract the bubble. When the radius is small, vector gravity
effectively produces a large repulsive potential which forces the bubble to
expand. As a result, the bubble radius oscillates between two turning
points. The radius of the $4\times 10^{6}M_{\odot }$ axion bubble at the
center of the Milky Way oscillates with a period of $20$ mins between $1$ $%
R_{\odot }$ and $1$ astronomical unit \cite{Svid07}.

This prediction has important implication for capturing the first image of
the supermassive object at the center of the Milky Way with the Event
Horizon Telescope (EHT) \cite{Godd17}. Namely, the oscillating axion bubble
produces shadow by bending light rays from the background sources only
during short time intervals when bubble size is smaller or of the order of
the gravitational radius $r_{g}=2GM/c^{2}=17$ $R_{\odot }$. Since typical
EHT image collection time is several hours the time averaging yields a much
fainter image than that expected from a static black hole in general
relativity and, hence, in the time-averaged image the shadow should be
almost invisible.

One should mention that first image of the Milky Way center with ALMA at 3.5
mm wavelength has been reported recently (see Fig. 5 in \cite{Issa19}). The
resolution of the detection is only slightly greater than the size of the
black hole shadow. Still, decrease in the intensity of light toward the
center should be visible if the Galactic center harbors a black hole. But
the image gets brighter (not dimmer) closer to the center. This agrees with
the prediction of vector gravity about lack of black holes.

On the other hand, a much heavier axion bubble in M87 galaxy ($M=4\times
10^{9}M_{\odot }$ \cite{Wals13}) does not expand substantially during
oscillations and its size remains of the order of $r_{g}$. As a consequence,
the axion bubble in M87 produces shadow comparable to that of a static black
hole. Due to bubble oscillations the bubble shadow in M87 should vary on a
timescale of a few days.

Vector gravity also provides an explanation of the dark energy as the energy
of longitudinal gravitational field induced by the expansion of the universe
and yields, with no free parameters, the value of $\Omega _{\Lambda }$ which
agrees with the results of Planck collaboration \cite{Planck14} and recent
results of the Dark Energy Survey. Moreover, recent gravitational wave
polarization analysis of GW170817 \cite{Abbo17a} showed that data are
compatible with vector gravity but not with general relativity \cite%
{Svid18,Svid18a}.

Similarly to general relativity, vector gravity postulates that the
gravitational field is coupled to matter through a metric tensor $f_{ik}$
which is, however, not an independent variable but rather a functional of
the vector gravitational field. In particular, action for a point particle
with mass $m$ moving in the gravitational field reads%
\begin{equation}
S_{\text{matter}}=-mc\int \sqrt{f_{ik}dx^{i}dx^{k}},  \label{d1a}
\end{equation}%
where $c$ is the speed of light. Action (\ref{d1a}) has the same form as in
general relativity, however, the tensor gravitational field $g_{ik}$ of
general relativity is now replaced with the equivalent metric $f_{ik}$ ($%
f_{ik}$ is a tensor under general coordinate transformations).

It is convenient to represent the vector gravitational field in terms of a
unit vector $u_{k}$ and a scalar $\phi $ related to the field absolute
value. Then in the Cartesian coordinate system of the background Euclidean
space the equivalent metric reads \cite{Svid17}%
\begin{equation}
f_{ik}=-e^{-2\phi }\delta _{ik}+2\cosh (2\phi )u_{i}u_{k},  \label{met}
\end{equation}%
while metric $\tilde{f}^{ik}$ inverse to $f_{ik}$, defined as $\tilde{f}%
^{ik}f_{im}=\delta _{m}^{k}$, is%
\begin{equation}
\tilde{f}^{ik}=-e^{2\phi }\delta ^{ik}+2\cosh (2\phi )u^{i}u^{k},
\label{meti}
\end{equation}%
where $\delta _{ik}=$diag$(1,1,1,1)$, 
\begin{equation*}
u^{i}=\delta ^{ik}u_{k},\quad u_{k}u^{k}=1,
\end{equation*}%
and $i,k=0,1,2,3$.

The total action for the gravitational field and matter is given by 
\begin{equation}
S=S_{\text{gravity}}+S_{\text{matter}},  \label{d3a}
\end{equation}%
where $S_{\text{matter}}$ is the action of matter written in curvilinear
coordinates with the metric $f_{ik}$. The action for the gravitational field 
$S_{\text{gravity}}$ is obtained from the requirement that symmetries of $S_{%
\text{matter}}$ and $S_{\text{gravity}}$ are the same. This requirement
yields a unique answer for $S_{\text{gravity}}$ \cite{Svid17}%
\begin{equation*}
S_{\text{gravity}}=\frac{c^{3}}{8\pi G}\int d^{4}x\left[ \frac{\partial \phi 
}{\partial x^{i}}\frac{\partial \phi }{\partial x^{k}}\left( -\delta
^{ik}+\left( 1-3e^{-4\phi }\right) u^{i}u^{k}\right) \right.
\end{equation*}%
\begin{equation*}
+\cosh ^{2}(2\phi )\frac{\partial u_{i}}{\partial x^{k}}\frac{\partial u_{m}%
}{\partial x^{l}}\Big(\delta ^{im}\delta ^{kl}-\delta ^{il}\delta ^{km}-
\end{equation*}%
\begin{equation}
\left. \left( 1+e^{-4\phi }\right) \delta ^{im}u^{k}u^{l}\Big)+2\left(
1+e^{-4\phi }\right) \frac{\partial \phi }{\partial x^{i}}\frac{\partial
u_{m}}{\partial x^{k}}\delta ^{im}u^{k}\right] ,  \label{fa2}
\end{equation}%
where $G$ is the gravitational constant.

Variation of the total action (\ref{d3a}) with respect to $\phi $ and the
unit vector $u_{k}$ gives equations for the gravitational field (see
Appendix A). Field equations (\ref{sss1}) and the action (\ref{fa2}) are not
generally covariant. However, they are invariant under coordinate
transformations that leave the background Euclidean metric $\delta _{ik}$
intact. These transformations, in particular, include rotations of the form%
\begin{equation}
x^{0}\rightarrow \frac{x^{0}+\frac{V}{c}x^{1}}{\sqrt{1+V^{2}/c^{2}}},\quad
x^{1}\rightarrow \frac{x^{1}-\frac{V}{c}x^{0}}{\sqrt{1+V^{2}/c^{2}}},
\label{r1}
\end{equation}%
which are analogous to the Lorentz transformations.

In vector gravity, motion of particles in gravitational field is described
by the same equation as in general relativity

\begin{equation}
\frac{d^{2}x^{b}}{ds^{2}}=\frac{1}{2}\tilde{f}^{bl}\left[ \frac{\partial
f_{ik}}{\partial x^{l}}-\frac{\partial f_{lk}}{\partial x^{i}}-\frac{%
\partial f_{il}}{\partial x^{k}}\right] \frac{dx^{i}}{ds}\frac{dx^{k}}{ds},
\label{r10aa}
\end{equation}%
where $ds=\sqrt{f_{ik}dx^{i}dx^{k}}$. In Eq. (\ref{r10aa}) the metric $%
g_{ik} $ of general relativity is replaced with the equivalent metric $%
f_{ik} $ and particles move along geodesics of $f_{ik}$.

One should emphasize that vector gravitational field lives in the
four-dimensional Euclidean manifold and raising and lowering of indexes in
the gravitational field action (\ref{fa2}) and equations (\ref{sss1}) is
carried out using the Euclidean metric $\delta _{ik}=$diag$(1,1,1,1)$.
However, all non-gravitational fields and matter sense the equivalent metric 
$f_{ik}$, that is geometry is effectively altered by the vector
gravitational field. This is why $f_{ik}$, rather than $\delta _{ik}$,
appears in the equation of motion (\ref{r10aa}). The equivalent metric $%
f_{ik}$ constitutes a manifold describing interaction with the gravitational
field. To avoid confusion between the two manifolds, we use tilde to denote
quantities obtained by raising of indexes using $f_{ik}$. For example, $%
\tilde{f}^{ik}$ stands for the metric inverse to $f_{ik}$.

According to vector gravity, transition between Euclidean geometry of the
equivalent metric and geometry of Minkowski signature occurred at the moment
of Big Bang. Before the Big Bang the vector gravitational field had no
preferred direction and was undergoing quantum fluctuations. The local
geometry in the pre-Big Bang era has Minkowski character and local direction
of the vector field determines the time-like dimension and the equivalent
metric (\ref{met}). As shown in \cite{Svid17}, the longitudinal component of
the vector gravitational field is not quantized and remains classical.
Hence, gravitational field does not undergo quantum fluctuations along the
field direction and time is a classical object. As a result, evolution of a
quantum system in time is well-defined in vector gravity. Namely, the time
derivative in the Heisenberg equation of motion has a meaning of the
derivative along the direction of the vector gravitational field.

Local Minkowski geometry allows for the field fluctuations which occur on a
Planck scale. Averaging over a small four-dimensional volume with size much
larger than Planck length and assuming that fluctuations are isotropic
yields 
\begin{equation*}
\left\langle u_{k}\right\rangle =0,\quad \left\langle
u_{i}u_{k}\right\rangle =\frac{1}{4}\delta _{ik},
\end{equation*}%
\begin{equation}
\left\langle f_{ik}\right\rangle =\frac{e^{-2\phi }}{4}\left( e^{4\phi
}-3\right) \delta _{ik}.  \label{b1}
\end{equation}%
That is before the Big Bang the equivalent metric has Euclidean character on
\textquotedblleft macroscopic\textquotedblright\ scales much larger than the
Planck length.

Big Bang is the point of quantum phase transition at which the gravitational
field vector acquires nonzero expectation value on the \textquotedblleft
macroscopic\textquotedblright\ scales $\left\langle u_{k}\right\rangle \neq
0 $. This expectation value serves as a transition order parameter. We
choose coordinate axis $x_{0}$ along the direction of $\left\langle
u_{k}\right\rangle $. In the disordered phase of universe the spatial
average of $u_{0}^{2}$ is 
\begin{equation*}
\left\langle u_{0}^{2}\right\rangle =\frac{1}{4}.
\end{equation*}%
Deviation of $\left\langle u_{0}^{2}\right\rangle $ from $1/4$ caused by
fluctuations can result in the signature flip of $\left\langle
f_{ik}\right\rangle $. Amplitude of the fluctuation which produces the
signature flip depends on the local value of $\phi $. Namely, spatial
averaging of Eq. (\ref{met}) yields that if $e^{4\phi }<3$ the signature
flip (phase transition) occurs if 
\begin{equation*}
\left\langle u_{0}^{2}\right\rangle =\frac{1}{1+e^{4\phi }}>\frac{1}{4}.
\end{equation*}%
At this point $\left\langle f_{00}\right\rangle $ changes sign from negative
to positive. For $e^{4\phi }>3$ the signature flip occurs at 
\begin{equation*}
\left\langle u_{0}^{2}\right\rangle =1-\frac{3}{1+e^{4\phi }}>\frac{1}{4}.
\end{equation*}%
At this point $\left\langle f_{\alpha \alpha }\right\rangle $ ($\alpha
=1,2,3 $) change sign from positive to negative.

According to Eq. (\ref{b1}), for 
\begin{equation*}
e^{4\phi }=3
\end{equation*}%
the average equivalent metric vanishes before the Big Bang $\left\langle
f_{ik}\right\rangle =0$ and the signature flip occurs when $\left\langle
u_{0}^{2}\right\rangle $ only slightly deviates from the value in the
disordered phase of universe 
\begin{equation*}
\left\langle u_{0}^{2}\right\rangle =\frac{1}{4}+\Delta ,
\end{equation*}%
where $\Delta $ is a small positive number. This deviation creates a nonzero
average equivalent metric with Minkowski character 
\begin{equation}
\left\langle f_{ik}\right\rangle =\frac{2\Delta }{\sqrt{3}}\text{diag}\left(
1,-\frac{1}{3},-\frac{1}{3},-\frac{1}{3}\right) .
\end{equation}

In the pre-Big Bang era $\phi $ is inhomogeneous in the four-dimensional
space. Big Bangs occur at points where $e^{4\phi }=3$. At such points a
small ordering of $u_{k}$ caused by fluctuations produces average equivalent
metric with Minkowski character yielding instability toward generation of
matter and gravitational waves and onset of the inflation stage \cite{Svid17}%
. Our universe began from one of such points. Subsequent universe expansion
resulted in exponentially large deviation of $\phi $ from the initial value
such that shortly after Big Bang $e^{-\phi }\ggg 1$. If we disregard
exponentially small terms $e^{\phi }$ compared to the exponentially large
terms of the order of $e^{-\phi }$ then gravitational field is no longer
\textquotedblleft absolute\textquotedblright . Namely, shift of $\phi $ by a
constant is equivalent to rescaling of coordinates.

In vector gravity the gravitational field is not coupled to itself through
the equivalent metric and \textquotedblleft feels\textquotedblright\ the
background geometry. This is a typical feature of metric theories of gravity
with a prior geometry \cite{Will93}. As a consequence, equations for the
vector gravitational field contain the background geometry $\delta _{ik}$,
while equations of matter motion contain only the equivalent metric $f_{ik}$%
. Despite this, in vector gravity gravitational waves travel with the speed
of light \cite{Svid17}. This is usually not the case in other alternative
theories of gravity. After the multi-messenging detection of the GW170817
coalescence of neutron stars \cite{Abbo17a}, where light and gravitational
waves were measured to travel at the same speed with an error of $10^{-15}$,
many alternative theories of gravity were excluded \cite{Saks17}.

In 1965 Weinberg found within a perturbative dynamical framework that
Maxwell's theory of electromagnetism and Einstein's theory of gravity are
essentially the unique Lorentz invariant theories of massless particles with
spin $1$ and $2$ respectively \cite{Wein65}. In vector gravity, the graviton
is a spin$-1$ massless particle. It was shown in \cite{Svid17} that
quantization of linearized equations of vector gravity yields a theory
equivalent to QED, which agrees with the Weinberg's findings.

One should also mention that the Weinberg-Witten theorem \cite{Wein80}
stating that massless particles with spin $j>1/2$ can not carry a
Lorentz-covariant current, while massless particles with spin $j>1$ cannot
carry a Lorentz-covariant stress-energy does not apply to vector gravity.
The reason is the same why Weinberg-Witten theorem does not forbid photons,
namely, both spin$-1$ photons in QED and spin$-1$ gravitons in vector
gravity carry no conserved charge.

It is remarkable that field equations of vector gravity (\ref{sss1}) can be
solved analytically for arbitrary static mass distribution \cite{Svid17}.
Namely, if point masses are located at $\mathbf{r}_{1}$, $\mathbf{r}_{2}$,
\ldots\ $\mathbf{r}_{N}$ then exact solution of the field equations for the
equivalent metric is 
\begin{equation}
f_{ik}=\left( 
\begin{array}{cccc}
e^{2\phi } & 0 & 0 & 0 \\ 
0 & -e^{-2\phi } & 0 & 0 \\ 
0 & 0 & -e^{-2\phi } & 0 \\ 
0 & 0 & 0 & -e^{-2\phi }%
\end{array}%
\right) ,  \label{x4}
\end{equation}%
where 
\begin{equation}
\phi (\mathbf{r})=-\frac{m_{1}}{|\mathbf{r}-\mathbf{r}_{1}|}-\ldots -\frac{%
m_{N}}{|\mathbf{r}-\mathbf{r}_{N}|}  \label{x5}
\end{equation}%
and $m_{l}$ ($l=1,\ldots ,N$) are constants determined by the value of
masses. Solution (\ref{x4}) shows lack of black holes in vector gravity.

However, in a general case, field equations (\ref{sss1}) are complicated.
The main purpose of the present paper is to simplify gravitational field
equations (\ref{sss1}) and make them more suitable for analytical and
numerical analyses. In addition, section \ref{dark} of this paper provides
new insights into dark energy.

\section{Simplified equations for gravitational field}

Due to expansion of the universe the spatial scale $a=e^{-\phi }$ has been
magnified in an exponentially large factor. Thus, shortly after Big Bang $%
e^{-\phi }$ became an exponentially large number ($e^{-\phi }\ggg 1$).
Therefore, one can disregard exponentially small terms $e^{\phi }$ compared
to the exponentially large terms of the order of $e^{-\phi }$.

As a result of cosmological expansion the gravitational field became
approximately uniform in the entire universe. We denote the coordinate along
the average direction of the gravitational field as $x^{0}=ct$. It
determines the cosmological reference frame. With the exponential accuracy
one can take $u_{0}\approx 1$, while $u^{\alpha }$ is of the order of $%
e^{2\phi }$.

Taking into account that $e^{-\phi }\ggg 1$ the equivalent metric (\ref{met}%
) in the cosmological reference frame reduces to a simple expression 
\begin{equation}
f_{00}=e^{2\phi }-e^{-2\phi }u^{2},\quad f_{0\alpha }=e^{-2\phi }u_{\alpha
},\quad f_{\alpha \beta }=-e^{-2\phi }\delta _{\alpha \beta },  \label{www1}
\end{equation}%
where $u^{2}=\delta _{\alpha \beta }u^{\alpha }u^{\beta }$ ($\alpha ,\beta
=1,2,3$). Spatial components of the metric (\ref{www1}) are diagonal. The
square of the interval reads%
\begin{equation*}
ds^{2}=e^{2\phi }(dx^{0})^{2}-e^{-2\phi }\left( d\mathbf{r}-\mathbf{u}%
dx^{0}\right) ^{2},
\end{equation*}%
where $\mathbf{u}=u^{\alpha }$ is a three dimensional vector which is
analogous to the vector potential in classical electrodynamics. Scalar $\phi 
$ and the three dimensional vector $\mathbf{u}$ are independent variables
that describe gravitational field (equivalent metric $f_{ik}$).

In metric (\ref{www1}) the action for a point particle (\ref{d1a}) reduces to%
\begin{equation}
S_{\text{matter}}=-mc^{2}\int dt\sqrt{e^{2\phi }-e^{-2\phi }\left( \frac{%
\mathbf{V}}{c}-\mathbf{u}\right) ^{2}},  \label{d1}
\end{equation}%
where $\mathbf{V}=d\mathbf{r}/dt$ is the particle velocity. Gravitational
field vector $\mathbf{u}$ shifts the particle velocity $\mathbf{V}$ by $c%
\mathbf{u}$. Particle Lagrangian%
\begin{equation*}
L=-mc^{2}\sqrt{e^{2\phi }-e^{-2\phi }\left( \frac{\mathbf{V}}{c}-\mathbf{u}%
\right) ^{2}}
\end{equation*}%
gives the following expression for the particle generalized momentum%
\begin{equation}
\mathbf{p}=\frac{\partial L}{\partial \mathbf{V}}=\frac{e^{-3\phi }m(\mathbf{%
V-}c\mathbf{u)}}{\sqrt{1-\left( \frac{\mathbf{V}}{c}-\mathbf{u}\right)
^{2}e^{-4\phi }}}.  \label{p1}
\end{equation}%
Equation (\ref{p1}) shows that velocity of a massive particle has a limiting
value determined by 
\begin{equation}
\left\vert \frac{\mathbf{V}}{c}-\mathbf{u}\right\vert =e^{2\phi }.
\label{b2}
\end{equation}%
This equation coincides with the equation for the velocity of light $\mathbf{%
V}$ propagating in gravitational field. If $\mathbf{u}\neq 0$ then speed of
light depends on the propagation direction $\mathbf{V}$.

Taking into account that $e^{-\phi }\ggg 1$ the gravitational field action (%
\ref{fa2}) in the cosmological reference frame reduces to%
\begin{equation*}
S_{\text{gravity}}=\frac{c^{3}}{8\pi G}\int d^{4}x\left[ -(\nabla \phi )^{2}+%
\frac{e^{-4\phi }}{4}\mathop{\rm curl^{2}}\mathbf{u}\right.
\end{equation*}%
\begin{equation}
\left. -3e^{-4\phi }\left( D_{t}\phi \right) ^{2}+2e^{-4\phi }\nabla \phi
\cdot D_{t}\mathbf{u-}\frac{e^{-8\phi }}{4}\left( D_{t}\mathbf{u}\right) ^{2}%
\right] ,  \label{a1}
\end{equation}%
where%
\begin{equation}
D_{t}=\frac{1}{c}\frac{\partial }{\partial t}+(\mathbf{u}\cdot \nabla )
\label{t1}
\end{equation}%
is the local time derivative. Recall that in vector gravity the time
coordinate is given by the direction of the four-vector $u^{k}$ which in the
present approximation reduces to $u^{k}=(1,\mathbf{u})$. The local time
derivative (\ref{t1}) is the derivative along the four-vector $u^{k}$,
namely, $D_{t}=u^{k}\partial /\partial x^{k}$. On the other hand, in the
present approximation, derivatives in the directions perpendicular to $u^{k}$
reduce to combinations of $\partial /\partial x^{\alpha }$ ($\alpha =1,2,3$).

Action (\ref{a1}) is invariant under transformations%
\begin{equation}
\phi \rightarrow \phi +\phi _{0},\quad \mathbf{r}\rightarrow e^{\phi _{0}}%
\mathbf{r},\quad t\rightarrow e^{-\phi _{0}}t,\quad \mathbf{u}\rightarrow
e^{2\phi _{0}}\mathbf{u},  \label{tr1}
\end{equation}%
where $\phi _{0}$ is an arbitrary constant.

One can show that action (\ref{a1}), upto irrelevant surface term, can be
written as 
\begin{equation}
S_{\text{gravity}}=-\frac{c^{3}}{16\pi G}\int d^{4}x\left( \sqrt{-f}R+\frac{1%
}{2}e^{-8\phi }\left( D_{t}\mathbf{u}\right) ^{2}\right) ,  \label{aGR}
\end{equation}%
where $R$ is the Ricci scalar calculated from the equivalent metric $f_{ik}$%
, $f=\det (f_{ik})$ and $\sqrt{-f}=e^{-2\phi }$. The first term in Eq. (\ref%
{aGR}) is the Einstein-Hilbert action of general relativity in which GR
metric $g_{ik}$ is replaced with the equivalent metric $f_{ik}$.

Variation of the total action (\ref{d3a}), where $S_{\text{gravity}}$ is
given by Eq. (\ref{a1}), with respect to $\phi $ and $\mathbf{u}$ yields
equations for the gravitational field. Variation of $S_{\text{matter}}$ can
be calculated using formula \cite{Land95} 
\begin{equation*}
\delta S_{\text{matter}}=-\frac{1}{2c}\int d^{4}x\sqrt{-f}T^{ik}\delta
f_{ik},
\end{equation*}%
where $T^{ik}$ is the energy-momentum tensor of matter. Variation of the
action (\ref{d3a}) yields the following equations for the gravitational
field ($\phi $ and $\mathbf{u}$) in the cosmological reference frame%
\begin{equation*}
\Delta \phi +3e^{-4\phi }\left[ D_{t}+\mathop{\rm div}\mathbf{u}-2D_{t}\phi %
\right] D_{t}\phi -e^{-4\phi }\mathop{\rm div}(D_{t}\mathbf{u)}
\end{equation*}%
\begin{equation}
-\frac{1}{2}e^{-4\phi }\mathop{\rm curl^{2}}\mathbf{u+}e^{-8\phi }\left(
D_{t}\mathbf{u}\right) ^{2}=\frac{8\pi G}{c^{4}}\left( T^{00}-\frac{%
e^{-2\phi }}{2}T\right) ,  \label{s1}
\end{equation}%
\begin{equation*}
\nabla \mathop{\rm div}\mathbf{u-}\Delta \mathbf{u}+4(\nabla \phi \cdot
\nabla )\mathbf{u}+4\left[ D_{t}\phi -\mathop{\rm div}\mathbf{u}-D_{t}\right]
\nabla \phi
\end{equation*}%
\begin{equation}
\mathbf{-}e^{-4\phi }\left[ \left( 8D_{t}\phi -\mathop{\rm div}\mathbf{u}%
-D_{t}\right) D_{t}\mathbf{u}+\nabla u^{\beta }\cdot D_{t}u_{\beta }\right] =%
\frac{16\pi G}{c^{4}}\mathbf{j},  \label{s2}
\end{equation}%
where $T=T^{mk}f_{mk}$ is the trace of the energy-momentum tensor and 
\begin{equation*}
j^{\alpha }=T^{0\alpha }-u^{\alpha }T^{00}.
\end{equation*}

Equations (\ref{s1}) and (\ref{s2}) are invariant under transformations (\ref%
{tr1}) which can be used to eliminate the cosmological background $\phi _{%
\text{cosm}}$. One can also obtain Eqs. (\ref{s1}) and (\ref{s2}) directly
from the gravitational field equations (\ref{sss1}) in the limit $e^{-\phi
}\ggg 1$. Namely, equation with $i=0$ yields Eq. (\ref{s1}). To obtain Eq. (%
\ref{s2}) one should take equation for $i=\alpha $ and subtract equation for 
$i=0$ multiplied by $u^{\alpha }$.

Equations (\ref{s1}) and (\ref{s2}) for the scalar $\phi $ and the three
dimensional vector $\mathbf{u}$ are the main equations of the vector theory
of gravity. They determine the equivalent metric which in the cosmological
Cartesian coordinates reads%
\begin{equation*}
f_{ik}=\left( 
\begin{array}{cccc}
e^{2\phi }-e^{-2\phi }u^{2} & e^{-2\phi }u^{x} & e^{-2\phi }u^{y} & 
e^{-2\phi }u^{z} \\ 
e^{-2\phi }u^{x} & -e^{-2\phi } & 0 & 0 \\ 
e^{-2\phi }u^{y} & 0 & -e^{-2\phi } & 0 \\ 
e^{-2\phi }u^{z} & 0 & 0 & -e^{-2\phi }%
\end{array}%
\right) .
\end{equation*}%
The inverse metric is given by%
\begin{equation*}
\tilde{f}^{00}=e^{-2\phi },\quad \tilde{f}^{0\alpha }=e^{-2\phi }u^{\alpha
},\quad \tilde{f}^{\alpha \beta }=e^{-2\phi }u^{\alpha }u^{\beta }-e^{2\phi
}\delta ^{\alpha \beta }.
\end{equation*}

Equation of motion of massive particles in gravitational field can be
obtained from Eq. (\ref{r10aa}) or directly from Lagrange's equation $\frac{d%
}{dt}\frac{\partial L}{\partial \mathbf{V}}=\frac{\partial L}{\partial 
\mathbf{r}}$ which yields%
\begin{equation}
\frac{d\mathbf{p}}{dt}=-\frac{mc^{2}\left( e^{2\phi }+e^{-2\phi }\left( 
\frac{\mathbf{V}}{c}-\mathbf{u}\right) ^{2}\right) }{\sqrt{1-\left( \frac{%
\mathbf{V}}{c}-\mathbf{u}\right) ^{2}e^{-4\phi }}}\nabla \phi -cp^{\beta
}\nabla u_{\beta },  \label{u11}
\end{equation}%
where $\mathbf{p}$ is the particle generalized momentum (\ref{p1}) and $%
d/dt=\partial /\partial t+\mathbf{V}\cdot \nabla $ is the total time
derivative.

One should remember, however, that Eqs. (\ref{s1}) and (\ref{s2}) do not
describe phenomena related to gravitational waves. Equations for the
radiative part of the gravitational field (which is quantized) depend on the
vacuum state of the field. Recall that vector gravity assumes that quantum
of gravitational field (graviton) is a composite particle assembled from
fermion-antifermion pairs \cite{Svid17}. If fermion states are empty (moment
of the Big Bang) then radiative part of the field is described by Eqs. (\ref%
{s1}) and (\ref{s2}). However, this vacuum is unstable toward generation of
matter and filling the fermion states. Such instability is the mechanism of
matter generation at the Big Bang. Shortly after the Big Bang the fermion
states become filled and matter generation comes to an end. For the filled
vacuum emission of a gravitational wave corresponds to creation of
fermion-antifermion hole pairs out of the filled fermion states. In this
case the radiative part of the field is described by Eq. (\ref{s2}) with the
opposite sign of $\mathbf{j}$. Namely, linearized equation for the
transverse radiative field ($\mathop{\rm
div}\mathbf{u=}$ $0$) far from the sources reads \cite{Svid17}%
\begin{equation}
\left( \frac{1}{c^{2}}\frac{\partial ^{2}}{\partial t^{2}}-\Delta \right) 
\mathbf{u}=-\frac{16\pi G}{c^{4}}\mathbf{j}_{tr},
\end{equation}%
where $\mathbf{j}_{tr}$ is the transverse part of $\mathbf{j}$. For the
filled vacuum the energy of gravitational waves is positive.

\section{Energy density for gravitational field}

If action of the system has the form 
\begin{equation*}
S=\frac{1}{c}\int d^{4}xL\left( A_{l},\frac{\partial A_{l}}{\partial x^{k}}%
\right) ,
\end{equation*}%
where the Lagrangian density $L$ is some function of the quantities $A_{l}$,
describing the state of the system, and of their first derivatives, then
energy density of the system $w$ can be calculated using formula \cite%
{Land95}

\begin{equation}
w=\sum_{l}\frac{\partial A_{l}}{\partial x^{0}}\frac{\partial L}{\partial 
\frac{\partial A_{l}}{\partial x^{0}}}-L.  \label{emt}
\end{equation}

Equation (\ref{a1}) yields 
\begin{equation*}
L_{\text{gravity}}=\frac{c^{4}}{8\pi G}\left[ -(\nabla \phi )^{2}+\frac{%
e^{-4\phi }}{4}\mathop{\rm curl^{2}}\mathbf{u}\right.
\end{equation*}%
\begin{equation*}
\left. -3e^{-4\phi }\left( D_{t}\phi \right) ^{2}+2e^{-4\phi }\nabla \phi
\cdot D_{t}\mathbf{u-}\frac{e^{-8\phi }}{4}\left( D_{t}\mathbf{u}\right) ^{2}%
\right] .
\end{equation*}%
Using Eq. (\ref{emt}), we obtain the following expression for the energy
density of gravitational field 
\begin{equation*}
w_{\text{field}}=\frac{c^{4}}{8\pi G}\left[ (\nabla \phi )^{2}-\frac{%
3e^{-4\phi }}{c^{2}}\left( \partial _{t}\phi \right) ^{2}\right.
\end{equation*}%
\begin{equation*}
-\frac{e^{-4\phi }}{4}\mathop{\rm curl^{2}}\mathbf{u-}\frac{e^{-8\phi }}{%
4c^{2}}\left( \partial _{t}\mathbf{u}\right) ^{2}-2e^{-4\phi }\nabla \phi
\cdot (\mathbf{u}\cdot \nabla )\mathbf{u}
\end{equation*}%
\begin{equation}
\left. +3e^{-4\phi }(\mathbf{u}\cdot \nabla \phi )^{2}+\frac{e^{-8\phi }}{4}%
\left( (\mathbf{u}\cdot \nabla )\mathbf{u}\right) ^{2}\right] .
\end{equation}%
This expression is valid for the vacuum of empty fermion states. For such
vacuum the energy of gravitational waves is negative which yields
instability toward generation of matter with positive energy and
gravitational waves with negative energy. As soon as fermion states are
filled the gravitational wave energy becomes positive. In particular, for
filled vacuum the energy density for a weak transverse gravitational wave is 
\cite{Svid17}%
\begin{equation*}
w_{\text{tr}}=\frac{c^{4}}{32\pi G}\left[ \frac{1}{c^{2}}\left( \frac{%
\partial \mathbf{u}}{\partial t}\right) ^{2}+\mathop{\rm curl^{2}}\mathbf{u}%
\right] .
\end{equation*}

For matter of density $\rho $ moving with velocity $\mathbf{V}$ the
Lagrangian density%
\begin{equation*}
L_{\text{matter}}=-\rho c^{2}\sqrt{e^{2\phi }-e^{-2\phi }\left( \frac{%
\mathbf{V}}{c}-\mathbf{u}\right) ^{2}}
\end{equation*}%
gives the following expression for the matter energy density%
\begin{equation}
w_{\text{matter}}=\rho c^{2}\frac{e^{2\phi }+e^{-2\phi }\mathbf{u}\left( 
\frac{\mathbf{V}}{c}-\mathbf{u}\right) }{\sqrt{e^{2\phi }-e^{-2\phi }\left( 
\frac{\mathbf{V}}{c}-\mathbf{u}\right) ^{2}}}.  \label{w1}
\end{equation}%
Equation (\ref{w1}) yields that upto terms quadratic in $\mathbf{u}$ and $%
\mathbf{V}$%
\begin{equation*}
w_{\text{matter}}=\rho c^{2}e^{\phi }+\frac{1}{2}\rho c^{2}e^{-3\phi }\left( 
\frac{V^{2}}{c^{2}}-u^{2}\right) .
\end{equation*}

\section{On the nature of dark energy}

\label{dark}

According to vector gravity, dark energy is the energy of longitudinal
gravitational field induced by the expansion of the universe \cite{Svid17}.
Universe expansion generates matter current which causes small deviations of
the four-vector gravitational field $A^{k}$ from the average cosmological
direction. These deviations yield nonzero cosmological constant $\Lambda $
in the universe evolution equation if the time axis is fixed by the
direction of $A^{k}$ at a certain moment $t_{0}$.

As shown in \cite{Svid17}, in vector gravity the equivalent metric 
\begin{equation}
f_{ik}=\left( 
\begin{array}{cccc}
\frac{1}{a^{2}} & 0 & 0 & 0 \\ 
0 & -a^{2} & 0 & 0 \\ 
0 & 0 & -a^{2} & 0 \\ 
0 & 0 & 0 & -a^{2}%
\end{array}%
\right)  \label{c2}
\end{equation}%
obeys evolution equation of the standard FLRW cosmology with the
cosmological term $\Lambda $. For example, for cold universe, equation for
the scaling factor $a(t)$ in the metric (\ref{c2}) reads%
\begin{equation}
-\frac{d^{2}}{dt^{2}}a^{2}(t)=\frac{16\pi G}{3}\left( \frac{\rho }{2a^{3}(t)}%
-\Lambda \right) ,  \label{qq1}
\end{equation}%
where $\rho $ and $\Lambda $ are independent of time ($\Lambda $ is usually
called $\rho _{\Lambda }$, the equivalent density). The constant $\rho $ has
a meaning of matter density for $a=1$. Eq. (\ref{qq1}) describes evolution
of the spatially averaged metric (\ref{c2}) which is uniform and isotropic
(depends only on time).

In vector gravity, as well as in general relativity, the cosmological term
is introduced into the action for the averaged metric using symmetry
arguments. Namely, symmetry arguments yield that spatial averaging can give
an effective gravitational field action with an additional cosmological term
of the form 
\begin{equation}
S_{\text{cosm}}=-c\Lambda \int d^{4}x\sqrt{-f},  \label{cosm}
\end{equation}%
where $\Lambda $ is a constant independent of the gravitational field and $%
f=\det (f_{ik})$. The cosmological term (\ref{cosm}) appears in vector
gravity due to replacement of the exact inhomogeneous equations with the
equations for the averaged metric which is spatially uniform and isotropic.
The value of $\Lambda $ can be obtained by matching the effective nonlinear
evolution equation (\ref{qq1}) with the exact linearized inhomogeneous
equations which do not contain $\Lambda $. Since gravitational field
equations in vector gravity are not generally covariant, the value of the
cosmological constant $\Lambda $ depends on a coordinate system.

In \cite{Svid17} we obtained $\Lambda $ in the reference frame of an
observer on Earth that takes a snapshot of the universe from a fixed point $%
O $ of the four-dimensional space at time $t_{0}$. In such frame, $t$ is a
coordinate of the Cartesian coordinate system in which the background
Euclidean metric is equal to $\delta _{ik}$ and the direction of the time
axis is determined by the location of the observer at time $t_{0}$. Namely,
the observer interprets the time axis as the instantaneous direction of the
gravitational field four-vector $A^{k}$ at the point $O$. Thus, location of
the observer at time $t_{0}$ fixes the time coordinate in the entire
four-dimensional space and from the perspective of the observer the universe
evolves along this time coordinate according to Eq. (\ref{qq1}) with nonzero 
$\Lambda $.

It is shown in \cite{Svid17} (by averaging the exact linearized
inhomogeneous equations without cosmological term) that in such coordinate
system in the vicinity of the observational point $O$ the scaling factor for
the cold universe satisfies condition

\begin{equation}
\frac{d^{2}}{dt^{2}}a^{2}(t_{0})=\frac{8\pi G\rho }{a^{3}(t_{0})},
\end{equation}%
which yields the following value of the cosmological constant%
\begin{equation}
\Lambda =\frac{2\rho }{a^{3}(t_{0})},  \label{s1a}
\end{equation}%
where $a(t_{0})$ is the value of the scaling factor at the space-time
position of the observer $O$.

Thus, vector gravity predicts that 
\begin{equation}
\frac{\Lambda }{\rho _{\text{critical}}(t_{0})}=\frac{2}{3},  \label{g1}
\end{equation}%
where $\rho _{\text{critical}}(t_{0})=\rho /a^{3}(t_{0})+\Lambda $ is the
critical density of the cold universe at the moment $t_{0}$ when the
observer takes a snapshot of the universe. Here we shed more light on this
result.

In vector gravity, gravitational field is a four-vector $A^{k}$ in a fixed
background four-dimensional Euclidean space $x^{i}$ ($i,k=0,1,2,3$). Let us
assume that we have solved the exact inhomogeneous field equations for the
whole universe and find $A^{k}(x^{i})$. The vector $A^{k}$ predominantly
points in the same direction everywhere, however there are deviations from
this direction that depend on $x^{i}$. These deviations, in particular, are
caused by the universe expansion itself.

In the cosmological model we approximate universe as homogeneous and
isotropic. In such a model the vector field $A^{k}(x^{i})$ is replaced by
its average over spatial coordinates. But what is the time coordinate and
what are the spatial coordinates? Recall that in the background Euclidean
space all coordinates are equivalent. At this stage we must specify what are
the spatial coordinates, that is specify the reference frame in which we
perform averaging.

Let us consider an observer located at a point $y^{i}$ in the
four-dimensional Euclidean space. Direction of $A^{k}$ at this point is the
time coordinate from the perspective of this observer. By making coordinate
transformation along the lines of Eq. (\ref{r1}) one can make the $x^{0}-$%
axis parallel to $A^{k}$ and denote $x^{0}=ct$, and $x^{1}$, $x^{2}$, $x^{3}$
as spatial coordinates $\mathbf{r}$. Thus, position of the observer fixes
the time and spatial coordinates in the whole Euclidean space. This is a
global Cartesian coordinate system associated with the point $y^{i}$.
Observer on Earth takes a snapshot of the whole universe in this global
coordinate system. Namely, the observer averages $A^{k}$ over $\mathbf{r}$
and obtains $\bar{A}^{k}(t)=\left\langle A^{k}(x^{i})\right\rangle _{\mathbf{%
r}}$ in this coordinate system.

If the observer is located at a different point $z^{i}$ the direction of $%
A^{k}$ at this point is different and, thus, division into time $t$ and
spatial coordinates $\mathbf{r}$ will not be the same. As a consequence, the
function $\bar{A}^{k}(t)$ will be different because it is obtained by
averaging of $A^{k}$ over different coordinates. Thus, equation for the
universe evolution in the cosmological model depends on the reference frame.

Such equation for the scaling factor $a(t)=e^{-\phi }$ was obtained in \cite%
{Svid17} using the averaging procedure outlined above. This procedure yields
uniform and isotropic equivalent metric (\ref{c2}) and for non relativistic
matter the evolution equation for $a(t)$ is given by Eq. (\ref{qq1}), or
after time integration 
\begin{equation}
\dot{a}^{2}(t)=\frac{8\pi G}{3}\left( \frac{\rho }{a^{3}(t)}+\Lambda \right)
,  \label{w5a}
\end{equation}%
where $\rho $ and $\Lambda $ are independent of time. However, the value of $%
\Lambda $ depends on the observer's reference frame. Local direction of $%
A^{k}$ at the observer's position determines the time coordinate $t$ and the
observer sees evolution of the universe $a(t)$ as a function of this time
coordinate.

Equation (\ref{s1a}) for $\Lambda $ has been obtained in \cite{Svid17} for
non-relativistic matter (present universe). Using similar procedure one can
show that for the radiation-dominated universe the answer will be%
\begin{equation*}
\Lambda =\frac{2\rho }{a^{4}(t_{0})},
\end{equation*}%
that is at time $t_{0}$ we also obtain 
\begin{equation}
\Omega _{\Lambda }(t_{0})=\frac{\Lambda }{\rho _{\text{critical}}(t_{0})}=%
\frac{2}{3}\approx 0.67,  \label{Lambda}
\end{equation}%
where for the radiation-dominated universe $\rho _{\text{critical}}(t)=\rho
/a^{4}(t)+\Lambda $.

The ratio (\ref{Lambda}) is independent of the universe equation of state
and on the time $t_{0}$ the observer measures $\Lambda $. However, the
critical density of the universe $\rho _{\text{critical}}(t)$ depends on
time. Eq. (\ref{Lambda}) holds only at the time $t_{0}$ for which direction
of the gravitational field four-vector $A^{k}$ coincides with the $t-$axis.
When the observer looks back in time by detecting light coming from distant
parts of the universe the direction of $A^{k}$ deviates from the $t-$axis at
the position of the light sources. As a result, at $t\neq t_{0}$ Eq. (\ref%
{Lambda}) is not satisfied.

Eq. (\ref{Lambda}) is a constraint obtained in vector gravity on the
evolution equation of the standard FLRW cosmology (\ref{qq1}) in the
reference frame of an observer that takes a snapshot of the universe at time 
$t_{0}$. $\Omega _{\Lambda }(t_{0})=2/3$ is a prediction of vector gravity
which is consistent with the experimental results of Planck collaboration 
\cite{Planck14} and the Dark Energy Survey $0.686\pm 0.02$.

Next we show that obtained results are also consistent with the Big Bang
nucleosynthesis (and other standard processes in the early universe such as
recombination, etc.). To this purpose we must consider evolution of the
scaling factor in a locally co-evolving reference frame, rather than in the
global inertial frame associated with the fixed Euclidean background. Dark
energy (universe expansion) changes direction of $A^{k}$ with time. If we
choose a local non-inertial coordinate system such that the time coordinate
always points along the direction of $A^{k}$ then this is a co-evolving
frame which is the proper frame to study Big Bang nucleosynthesis.

Next we show that in the co-evolving frame the cosmological constant in Eq. (%
\ref{qq1}) is equal to zero. Therefore, in vector gravity the Big Bang
nucleosynthesis proceeds the same way as in general relativity which yields
that contribution from the cosmological term is negligible for the early
universe. To be specific, we will assume non-relativistic matter, however,
the answer is valid for a general equation of state.

It is convenient to work with the equivalent metric linearized near the
Minkowski space-time. The linearized version of Eq. (\ref{qq1}) reads \cite%
{Svid17} 
\begin{equation}
3\left\langle \ddot{h}_{00}\right\rangle +16\pi G\Lambda =8\pi G\rho (t_{0}),
\label{s10}
\end{equation}%
where $\left\langle h_{00}\right\rangle $ is a component of the spatially
averaged metric 
\begin{equation*}
\left\langle f_{ik}\right\rangle =\left( 
\begin{array}{cccc}
1+\left\langle h_{00}\right\rangle & 0 & 0 & 0 \\ 
0 & -1+\left\langle h_{00}\right\rangle & 0 & 0 \\ 
0 & 0 & -1+\left\langle h_{00}\right\rangle & 0 \\ 
0 & 0 & 0 & -1+\left\langle h_{00}\right\rangle%
\end{array}%
\right)
\end{equation*}%
and $\rho (t_{0})=\rho /a^{3}(t_{0})$.

In Ref. \cite{Svid17} it has been shown that components of the linearized
spatially inhomogeneous metric before averaging 
\begin{equation}
f_{ik}=\left( 
\begin{array}{cccc}
1+h_{00} & h_{01} & h_{02} & h_{03} \\ 
h_{01} & -1+h_{00} & 0 & 0 \\ 
h_{02} & 0 & -1+h_{00} & 0 \\ 
h_{03} & 0 & 0 & -1+h_{00}%
\end{array}%
\right)  \label{s0}
\end{equation}%
obey equations 
\begin{equation}
\Delta h_{00}+3\frac{\partial ^{2}h_{00}}{\partial x^{0}\partial x^{0}}-2%
\frac{\partial ^{2}h_{0\beta }}{\partial x^{0}\partial x^{\beta }}=\frac{%
8\pi G}{c^{4}}T^{00},  \label{wf1}
\end{equation}%
\begin{equation}
\left( \frac{\partial ^{2}}{\partial x^{0}\partial x^{0}}-\Delta \right)
h_{0\alpha }+\frac{\partial ^{2}h_{0\beta }}{\partial x^{\alpha }\partial
x^{\beta }}-2\frac{\partial ^{2}h_{00}}{\partial x^{\alpha }\partial x^{0}}=%
\frac{16\pi G}{c^{4}}T^{\alpha 0}.  \label{wf2}
\end{equation}%
Using the continuity equation for matter%
\begin{equation*}
\frac{\partial T^{00}}{\partial x^{0}}+\frac{\partial T^{0\alpha }}{\partial
x^{\alpha }}=0
\end{equation*}%
one can reduce Eqs. (\ref{wf1}) and (\ref{wf2}) to \cite{Svid17}%
\begin{equation}
\frac{1}{c^{2}}\ddot{h}_{00}=\Delta h_{00}-\frac{8\pi G}{c^{2}}\rho (t,%
\mathbf{r}),  \label{cosx}
\end{equation}%
\begin{equation}
\frac{\partial ^{2}h_{0\alpha }}{\partial x^{0}\partial x^{\alpha }}=2\frac{%
\partial ^{2}h_{00}}{\partial x^{0}\partial x^{0}}.  \label{cos9}
\end{equation}%
Eq. (\ref{cos9}) has the following local solution valid in the linear
approximation 
\begin{equation}
h_{0\alpha }(t,\mathbf{r})=\frac{2}{3c}\ddot{h}_{00}(t_{0})(t-t_{0})x^{%
\alpha }.  \label{ev1}
\end{equation}%
The coordinates $x^{0}=ct$ and $x^{\alpha }$ ($\alpha =1,2,3$) determine the
global inertial reference frame in the Euclidean space. At the observer's
position ($x^{\alpha }=0$) in the linear approximation $h_{0\alpha }=0$.
That is the gravitational field $4-$vector points along the $x^{0}-$axis.

If we average $f_{ik}$ over the global spatial coordinates $x^{\alpha }$ we
find that $\left\langle f_{ik}\right\rangle $ is diagonal. Averaging Eq. (%
\ref{cosx}) yields

\begin{equation}
\left\langle \ddot{h}_{00}\right\rangle =-8\pi G\rho (t_{0}).  \label{s11}
\end{equation}%
Matching this with Eq. (\ref{s10}), we obtain $\Lambda =2\rho (t_{0})\neq 0$.

Next, we make transformation to the co-evolving coordinates in which
spatially nonuniform metric (\ref{s0}) is diagonal in the local region.
Recall that for an infinitesimal transformation of coordinates%
\begin{equation*}
x^{\prime k}=x^{k}+\xi ^{k}
\end{equation*}%
the metric transforms as \cite{Land95}%
\begin{equation*}
g_{ik}^{\prime }=g_{ik}-\xi _{i;k}-\xi _{k;i}.
\end{equation*}%
For the present case the covariant derivatives can be replaced with partial
derivatives.

Transformation of coordinates%
\begin{equation}
x^{\prime \alpha }=x^{\alpha }-\frac{1}{3}\ddot{h}%
_{00}(t_{0})(t-t_{0})^{2}x^{\alpha },  \label{tc1}
\end{equation}%
\begin{equation}
x^{\prime 0}=x^{0}+\frac{c}{9}\ddot{h}_{00}(t_{0})(t-t_{0})^{3},  \label{tc2}
\end{equation}%
yields a diagonal metric%
\begin{equation*}
f_{ik}^{\text{ }\prime }=\left( 
\begin{array}{cccc}
1+\tilde{h}_{00} & 0 & 0 & 0 \\ 
0 & -1+\tilde{h}_{00} & 0 & 0 \\ 
0 & 0 & -1+\tilde{h}_{00} & 0 \\ 
0 & 0 & 0 & -1+\tilde{h}_{00}%
\end{array}%
\right) ,
\end{equation*}%
where 
\begin{equation*}
\tilde{h}_{00}=h_{00}-\frac{2}{3}\ddot{h}_{00}(t_{0})(t-t_{0})^{2}.
\end{equation*}%
Taking the second order time derivative we find that $\tilde{h}_{00}$ obeys
differential equation%
\begin{equation*}
\frac{d^{2}\tilde{h}_{00}}{dt^{2}}=-\frac{1}{3}\ddot{h}_{00}(t_{0}).
\end{equation*}%
Plug in here Eq. (\ref{s11}) gives 
\begin{equation*}
3\frac{d^{2}\tilde{h}_{00}}{dt^{2}}=8\pi G\rho (t_{0}),
\end{equation*}%
that is $\tilde{h}_{00}$ obeys Eq. (\ref{s10}) with $\Lambda =0$. So, in the
co-evolving frame $\Lambda =0$ and the universe expansion is decelerating.

\section{Summary}

In vector gravity universe evolution is described by the equation of the
standard FLRW cosmology with a cosmological term $\Lambda $. However,
contrary to general relativity, the value of the cosmological constant $%
\Lambda $ in vector gravity depends on a coordinate system.

According to vector gravity, the Euclidean geometry of the universe is
altered by the vector gravitational field $A^{k}$. The direction of $A^{k}$
gives the time coordinate, while perpendicular directions are spatial
coordinates. Universe expansion yields change of the direction of $A^{k}$
with time. This change appears as the cosmological constant (dark energy) in
the evolution equation if we look at the universe globally in the fixed
inertial reference frame of the background Euclidean space. This is what an
observer on Earth does when he takes a snapshot of the universe tacitly
assuming that the time coordinate was always pointing in the same direction
given by the direction of $A^{k}$ at the moment of observation $t_{0}$.

In this special coordinate system associated with the observer on Earth the $%
t-$axis points in the direction of $A^{k}$ only at time $t_{0}$. When the
observer looks back in time by detecting light coming from distant parts of
the universe the direction of $A^{k}$ deviates from the $t-$axis at the
position of the light sources. As a result, in this coordinate system the
universe evolves as if there is a nonzero cosmological constant $\Lambda
=2\rho _{\text{critical}}(t_{0})/3$, where $\rho _{\text{critical}}(t_{0})$
is the critical density of the universe at the moment $t_{0}$.

However, in the local co-evolving reference frame, in which the time
coordinate is evolving together with the universe always aiming in the
direction of $A^{k}$, the dark energy produces no effect on the universe
evolution, and, as a result, it does not alter the Big Bang nucleosynthesis
and galaxy formation.

Relation between Hubble parameters $H$ in the fix and co-evolving frames can
be obtained by applying the coordinate transformation (\ref{tc1}) and (\ref%
{tc2}), which yields that in the vicinity of $t_{0}$%
\begin{equation}
H^{\prime }=H-\frac{2}{3}\ddot{h}_{00}(t_{0})(t-t_{0}).  \label{tc3}
\end{equation}%
Eq. (\ref{tc3}) shows that rate of change of Hubble parameter is different
in the two frames. Thus, universe expansion can be accelerating in one
frame, and decelerating in the other. However, in both frames the spatially
averaged metric has the same form given by Eq. (\ref{c2}).

Present analysis also answers the question about the fate of the universe
which is determined by the evolution of the scaling factor in the
co-evolving frame. Namely, the universe will expand forever at a continually
decelerating rate, with expansion asymptotically approaching zero. This is
what is expected for spatially flat universe in absence of exotic forms of
energy.

Exponential expansion of the universe at the moment of Big Bang allows us
substantially simplify equations for the gravitational field and reduce them
to a simple form (\ref{s1}), (\ref{s2}). Apart from being much simpler,
these equations have more symmetries than original field equations (\ref%
{sss1}). In particular, Eqs. (\ref{s1}) and (\ref{s2}) are invariant under
transformations (\ref{tr1}).

Simplified gravitational field equations (\ref{s1}) and (\ref{s2}) open a
perspective for a rapid development of the vector theory of gravity and
expand the class of problems for which analytical solutions can be obtained.
E.g., one can use them to find, in the framework of vector gravity,
stationary metric produced by a spinning mass in cylindrical or spherical
geometries.

\vspace{0.3cm}

\begin{acknowledgments}
This work was supported by the Air Force Office of Scientific Research
(Award No. FA9550-18-1-0141), the Office of Naval Research (Award Nos.
N00014-16-1-3054 and N00014-16-1-2578) and the Robert A. Welch Foundation
(Award A-1261).
\end{acknowledgments}

\appendix

\section{Gravitational field equations}

Equations for gravitational field in the background Euclidean space read 
\cite{Svid17}

\begin{equation*}
\left[ \delta ^{mk}u^{i}-2\delta ^{im}u^{k}+\left( 1+3e^{-4\phi }\right)
u^{m}u^{k}u^{i}\right] \frac{\partial ^{2}\phi }{\partial x^{m}\partial x^{k}%
}
\end{equation*}
\begin{equation*}
+2\left[ \delta ^{im}-\left( 3e^{-4\phi }+1\right) u^{m}u^{i}\right] \frac{%
\partial \phi }{\partial x^{m}}\frac{\partial \phi }{\partial x^{k}}u^{k}
\end{equation*}
\begin{equation*}
+2\left[ e^{4\phi }\left( \delta _{l}^{k}\delta ^{im}-\delta _{l}^{i}\delta
^{mk}\right) +\delta _{l}^{i}\delta ^{mk}-\delta _{l}^{m}\delta ^{ik}\right] 
\frac{\partial \phi }{\partial x^{k}}\frac{\partial u^{l}}{\partial x^{m}}
\end{equation*}
\begin{equation*}
+\left[ 2\left( e^{4\phi }-2e^{-4\phi }-1\right) \delta
_{l}^{i}u^{m}u^{k}-\left( 1-3e^{-4\phi }\right) \delta
_{l}^{m}u^{i}u^{k}\right.
\end{equation*}

\begin{equation*}
\left. -\left( 2e^{4\phi }-3e^{-4\phi }+1\right) \delta _{l}^{k}u^{m}u^{i} 
\right] \frac{\partial \phi }{\partial x^{k}}\frac{\partial u^{l}}{\partial
x^{m}}
\end{equation*}
\begin{equation*}
+\cosh (2\phi )\left[ e^{2\phi }\frac{\partial }{\partial x^{k}}\left( \frac{%
\partial u^{k}}{\partial x_{i}}-\frac{\partial u^{i}}{\partial x_{k}}\right)
+e^{-2\phi }u_{m}u^{i}\frac{\partial ^{2}u^{m}}{\partial x_{k}\partial x^{k}}%
\right.
\end{equation*}
\begin{equation*}
+\left. 2\cosh (2\phi )u^{k}u^{l}\frac{\partial ^{2}u^{i}}{\partial
x^{l}\partial x^{k}}-\left( e^{2\phi }+2e^{-2\phi }\right) u^{m}u^{i}\frac{%
\partial ^{2}u^{k}}{\partial x^{k}\partial x^{m}}\right]
\end{equation*}
\begin{equation*}
+2\cosh ^{2}(2\phi )\left[ \frac{\partial u^{i}}{\partial x^{k}}\frac{%
\partial }{\partial x^{m}}\left( u^{k}u^{m}\right) -\frac{\partial u_{k}}{%
\partial x_{i}}\frac{\partial u^{k}}{\partial x^{l}}u^{l}\right.
\end{equation*}
\begin{equation*}
\left. -\frac{\partial u^{k}}{\partial x^{m}}\frac{\partial u^{m}}{\partial
x^{k}}u^{i}+\left( 1+2e^{-4\phi }\right) \frac{\partial u^{k}}{\partial x^{m}%
}\frac{\partial u_{k}}{\partial x^{l}}u^{m}u^{l}u^{i}\right]
\end{equation*}
\begin{equation}
=\frac{8\pi G}{c^{4}}\left( T^{ik}-\frac{T}{2}\tilde{f}^{ik}\right) u_{k},
\label{sss1}
\end{equation}
where $T^{ik}$ is the energy-momentum tensor of matter and $T=T^{mk}f_{mk}$
is the trace of the energy-momentum tensor. Equations (\ref{sss1}) for $\phi 
$ and $u_{k}$ are written in Euclidean metric which means that raising and
lowering of indexes is carried out using $\delta _{ik}=$diag$(1,1,1,1)$.

\end{document}